\documentstyle[12pt,epsf]{article}

\title{Motion of a test body in the presence of an external scalar field which
respects the weak equivalence principle}

\author{J.  P.  Mbelek \\ Service d'Astrophysique, C.E.  Saclay \\ F-91191
Gif-sur-Yvette Cedex, France}

\begin{document} \maketitle \baselineskip=8mm

\begin{abstract} It is shown that the main contribution to the rotational curve
of a spiral galaxy may be due essentially to the interaction, in the general
relativistic spacetime, of the galactic matter with a very light long range
scalar field which respects the weak equivalence principle.  The comparison of
the theoretical results with 23 spiral galaxy rotation curves shows a good
agreement between our proposal and observations.  \end{abstract}

\section{Introduction} According to newtonian dynamics and Newton's
inverse-square force law of gravitation, the circular velocity around an
isolated body should decrease with the distance, $r$, to the centre like
$\frac{1}{\sqrt{r}}$.  However, observations show that the rotation curves of
many spiral galaxies flatten at large distances.  The current interpretation
consists to invoke the existence of dark matter though its nature and
distribution in space are not yet clearly defined~\cite{1}.

Alternative proposals, which provide explanations for observed galactic
rotation curves without the need of dark matter, have been developed by several
authors.  In that line of thinking, the Modified Newtonian Dynamics (MOND)
proposed by Milgrom~\cite{2} introduces a "universal" constant $a_{0}$ of the
dimension of an acceleration such that newtonian dynamics breaks down at
acceleration much smaller than $a_{0}$ (or equivalently Newton's law of
gravitation fails when the magnitude of the potential gradient is much smaller
than $a_{0}$).  This acceleration constant turns out to be of the order of
$cH_{0}$ suggesting a cosmological link~\cite{3}.  However, actually the
equality $a_{0} = cH_{0}$ has no theoretical basis in the framework of MOND and
it turns out that at the cosmological level MOND fails to satisfy the
cosmological principle~\cite{4}.  Besides, while having some successes at the
scale of galaxies~\cite{5, 6, 7} and clusters of galaxies~\cite{8, 9}, the MOND
fails in the laboratory\footnote{This conclusion is rejected by M.  Milgrom,
since the experiment has been performed on earth where the gravitational
pulling is much greater than $a_{0}$.}~\cite{10}.  Moreover, at the scale of
clusters of galaxies, Gerbal {\sl et al.}  have found that a certain amount of
dark matter is required even with MOND~\cite{11, 12}.

In the framework of the conformal Weyl gravity, Mannheim and Kazanas obtained a
complete, exact exterior solution for a static, spherically symmetric
source~\cite{13}.  In addition to the exterior Schwarzschild solution, their
solution contains an extra gravitational potential term $\gamma r$ which grows
linearly with the distance $r$ to the centre.  A cosmological link is also made
by these authors who noted intringuingly that their parameter $\gamma$ is
roughly the value of the inverse of the Hubble length, $\frac{c}{H_{0}}$.
Their solution is able to interpolate between those of Robertson-Walker and
Schwarzchild in a continuous and smooth manner by exploiting the conformal
structure of the conformal Weyl gravity.  This solution has been applied
successfully by Mannheim to the galactic rotation curves of spiral
galaxies~\cite{14}.  However, by matching the exterior solution to an interior
one that satisfies the weak energy condition and a regularity condition at the
centre, Perlick and Xu~\cite{15} have shown that this leads to contradiction of
Mannheim and Kazanas's suggestion.  They conclude that the conformal Weyl
gravity is not able to give a viable model of the solar system.  Besides, it
seems that the cosmological models derived from the conformal Weyl gravity
fails to fulfill simultaneously the observational constraints on present
cosmological parameters and on primordial light element abundances~\cite{16}.

Other authors have suggested that Newton's law of gravity, which describes so
well the motion of moons and planets in the solar system, may break down over
distances comparable to the size of a galaxy.  So, alternative force laws have
been proposed~\cite{17, 18, 19}.  However, one should bear in mind that
newtonian mechanics is the only appropriate approximation of general relativity
in the weak field and low velocity limit (see the criticism of D.
Lindley,~\cite{31}).

In this paper, we propose a new dark matter scheme in the framework of
Einstein's general relativity and its weak field and low velocity limit,
newtonian gravity.  Our new dark matter candidate is a long range neutral
massive scalar field directly coupled to matter, unlike the Brans-Dicke theory
in which the scalar field does not exert any direct influence on matter (its
only role is that of participant in the field equations that determine the
geometry of spacetime)~\cite{20}.  In addition, one requires that the scalar
field under consideration, $\phi$, respects the weak equivalence principle.
Let us notice that, if one takes into account the gravitational coupling of
$\phi$, with the ordinary matter, then the $g_{\alpha\beta}$'$^{s}$ will depend
not only on the $x^{\mu}$'$^{s}$ but also on $\phi$, that is $g_{\alpha\beta} =
g_{\alpha\beta}(x^{\mu}, \phi)$.  The plan of this paper is as follows.  In
section 2, we establish the equation of motion of a test body in the presence
of the light scalar field, $\phi$.  In section 3, we determine the potential
$V_{\phi}$ that couples the ordinary matter to the scalar field.  In section 4,
the results of both sections 2 and 3 are combined and applied to the rotational
curves of spiral galaxies.  Finally, in section 5, twenty three rotation curves
are confronted with the results of the previous sections.

\section{The equation of motion in the presence of a long range scalar field}
In the absence of any external field, the equation of motion of a test body
writes in general relativity \begin{equation} \label{1} (u^{\alpha}
\nabla_{\alpha}) u^{\mu} = 0 \end{equation} where $u_{\alpha} =
\frac{dx_{\alpha}}{ds}$ denotes the velocity four-vector of the test body and
$ds^{2} = g_{\alpha\beta} dx^{\alpha} dx^{\beta}$ is the metric and
$\nabla_{\alpha}$ is the covariant derivatives ($\alpha$, $\beta = 0, 1, 2,
3$).  In the same manner as in the presence of an electromagnetic field, in the
presence of a scalar field, $\phi$, a force term enters in the right-hand side
of equation (\ref{1}).  This force term would be {\sl a priori} of the simple
form $- \frac{1}{c^{2}} \frac{\partial V_{\phi}}{\partial x^{\alpha}}$ involved
by the lagrangian $L = - \frac{1}{2} g_{\alpha\beta} u^{\alpha} u^{\beta} -
\frac{V_{\phi}}{c^{2}}$, where the potential $V_{\phi}$ depends on $\phi$
($V_{\phi} = 0$ if $\phi = 0$) and is assumed to be the same for all matter at
any point of spacetime to ensure obedience to the weak equivalence principle.
Thus one would write

\begin{equation} \label{2} (u^{\alpha} \nabla_{\alpha}) u^{\mu} = -
\frac{1}{c^{2}} \frac{\partial V_{\phi}}{\partial x^{\mu}}.  \end{equation}
However, equation (\ref{2}) is not satisfactory, since the unitarity condition
$u^{\mu} u_{\mu} = 1$ implies

\begin{equation} \label{3} u^{\mu} (u^{\alpha} \nabla_{\alpha}) u_{\mu} = 0.
\end{equation} In order to satisfy relation (\ref{3}), there should be at least
the additional term $\frac{1}{c^{2}} \frac{dV_{\phi}}{ds} u_{\mu}$ to the
right-hand side of equation (\ref{2}).  Finally, the correct equation of motion
writes

\begin{equation} \label{4} (u^{\alpha} \nabla_{\alpha}) u^{\mu} = -
\frac{1}{c^{2}} \frac{\partial V_{\phi}}{\partial x^{\mu}} + \frac{1}{c^{2}}
\frac{dV_{\phi}}{ds} u_{\mu} \end{equation} or making appear explicitly the
Christoffel symbols,

\begin{equation} \label{5} \frac{du^{\mu}}{ds} + {\Gamma}^{\mu}_{\alpha\beta}
u^{\alpha} u^{\beta} = - \frac{1}{c^{2}} \partial^{\mu} V_{\phi} +
\frac{1}{c^{2}} \frac{dV_{\phi}}{ds} u^{\mu}.  \end{equation} One may derive
equation (\ref{4}, \ref{5}) from the lagrangian

\begin{equation} \label{6} L = - \frac{e^{- V_{\phi}/c^{2}}}{2}
(g_{\alpha\beta} u^{\alpha} u^{\beta} + 1).  \end{equation}

\subsection{Small speeds and weak fields approximation} For small speeds and
weak gravitational fields, with respect to a galilean referential, it comes
${\Gamma}^{\alpha}_{00} \approx - \frac{1}{2} \partial^{\alpha} g_{00}$.  Now,
$g_{\alpha\beta}(x^{\mu}, \phi) \approx g_{\alpha\beta}(x^{\mu}, 0) +
\left(\frac{\partial g_{\alpha\beta}}{\partial \phi} \right)_{\phi = 0} \phi$
since we are interested by the weak $\phi$-field approximation.  One may also
write $g_{\alpha\beta}(x^{\mu}, \phi) \approx g_{\alpha\beta}(x^{\mu}, 0) +
\left(\frac{\partial g_{\alpha\beta}}{\partial V_{\phi}} \right)_{V_{\phi} = 0}
V_{\phi}$, for $V_{\phi}$ is proportional to $\phi$ at the first order.
Moreover, for small speeds and weak gravitational fields $g_{00}(x^{\mu}, 0)
\approx 1 + 2 \frac{V_{N}}{c^{2}}$ whence $g_{00}(x^{\mu}, \phi) \approx 1 + 2
\frac{V_{N} + (1 + f) V_{\phi}}{c^{2}}$, where $V_{N}$ denotes the newtonian
gravitational potential and we have set $\left(\frac{\partial g_{00}}{\partial
V_{\phi}} \right)_{V_{\phi} = 0} = \frac{2}{c^{2}} (1 + f)$ (it will be found
in a next study on the mass distribution that $f$ is positive and typically in
the range $10^{-9} - 10^{-3}$ in spiral galaxies).  Consequently equation
(\ref{5}) reduces at the first order to

\begin{equation} \label{7} \frac{d^{2}\vec{r}}{dt^{2}} = - \vec{\nabla} (V_{N}
+ f V_{\phi}) + \frac{1}{c^{2}} \frac{dV_{\phi}}{dt} \frac{\vec{r}}{dt}
\end{equation} where $\frac{dV_{\phi}}{dt} = \frac{\partial V_{\phi}}{\partial
t} + \frac{\partial V_{\phi}}{\partial r} \frac{dr}{dt} + \frac{\partial
V_{\phi}}{\partial \theta} \frac{d\theta}{dt} + \frac{\partial
V_{\phi}}{\partial \varphi} \frac{d\varphi}{dt}$ in spherical coordinate $(r,
\theta, \varphi$).  Particularly, for planar orbits in plane polar coordinates
$(r, \theta)$, one gets from the above equation

\begin{equation} \label{8} \frac{dv_{r}}{dt} - \frac{v_{\theta}^{2}}{r} = -
\frac{\partial (V_{N} + fV_{\phi})}{\partial r} + \frac{1}{c^{2}}
\frac{dV_{\phi}}{dt} v_{r} \end{equation}

\begin{equation} \label{9} \frac{1}{r} \frac{d(v_{\theta}r)}{dt} = -
\frac{1}{r} \frac{\partial (V_{N} + fV_{\phi})}{\partial \theta} +
\frac{1}{c^{2}} \frac{dV_{\phi}}{dt} v_{\theta} \end{equation} where $v_{r} =
\frac{dr}{dt}$ is the radial velocity, $v_{\theta} = r \frac{d\theta}{dt}$ the
tangential velocity.

\subsection{Static spherically symmetric fields} For static spherically
symmetric fields, equations (\ref{8}) and (\ref{9}) become

\begin{equation} \label{10} \frac{dv_{r}}{dt} - \frac{v_{\theta}^{2}}{r} = - G
\frac{m(r)}{r^{2}} - f \frac{\partial V_{\phi}}{\partial r} \end{equation}

\begin{equation} \label{11} \frac{1}{r} \frac{d(v_{\theta}r)}{dt} =
\frac{1}{c^{2}} \frac{dV_{\phi}}{dt} v_{\theta} \end{equation} since $v_{r} \ll
c$, $\frac{\partial V_{\phi}}{\partial \theta} = \frac{\partial V_{N}}{\partial
\theta} = 0$ and $\frac{\partial V_{N}}{\partial r} = G \frac{m(r)}{r^{2}}$,
where $G$ is the gravitational constant and $m(r)$ denotes the mass up to
radius $r$.  On integrating through equation (\ref{11}), one gets

\begin{equation} \label{12} v_{\theta} = J
\frac{\exp{\left(\frac{V_{\phi}}{c^{2}}\right)}}{r}\end{equation} where $J$ is
a constant which would represent the angular momentum per unit mass if the
scalar field $\phi$ were not present.

\section{Derivation of the potential $V_{\phi}$} In as much as we are
interested by the weak field approximation only, the potential $V_{\phi}$ is
simply a linear function of the scalar field, $\phi$.  Now, the equation of a
scalar field, $\phi$, writes in the weak field approximation

\begin{equation} \label{13} \partial^{\mu} \partial_{\mu} \phi +
(\frac{m_{\phi}c}{\hbar})^{2} \phi = - V \phi.  \end{equation} The above
equation, as it is well known, follows from the lagrangian density of a real
scalar field $ L = \frac{1}{2} \partial^{\mu} \phi \,\partial_{\mu} \phi -
U(\phi)$, where $U(\phi) = \frac{1}{2} \,(\frac{m_{\phi}c}{\hbar})^{2}
\,\phi^{2} + \int V \phi \,d\phi$ denotes the potential energy of the scalar
field $\phi$ and $m_{\phi}$ is its mass.  Hereafter, we assume a
phenomenological effective potential energy of the form

\begin{equation} \label{14} U(\phi) = \frac{1}{2} (\frac{m_{\phi}c}{\hbar})^{2}
\phi^{2} + q \phi^{p} \end{equation} where $p$ and $q$ are real constants.  Of
course, whenever $p$ is an integer, $p \geq 3$.  In the following we determine
only the exponent $p$, in as much as the potential $U(\phi)$ is assumed
negligible with respect to the ordinary matter mass-energy.  Besides, since we
have considered a long range scalar field $\phi$, we may drop the mass term in
equation (\ref{13}) and consider the massless field equation

\begin{equation} \label{15} \partial^{\mu} \partial_{\mu} \phi = - V \phi.
\end{equation} Then, the static spherically symmetric solution $\phi = \phi
(r)$ satisfies the following equation

\begin{equation} \label{16} \frac{d^{2}\phi}{dr^{2}} + \frac{2}{r}
\frac{d\phi}{dr} = V \phi \end{equation} and the potential energy reduces to
the power law

\begin{equation} \label{17} U(\phi) = q \phi^{p}.  \end{equation} So, we look
for a solution of the form

\begin{equation} \label{18} \phi \propto r^{k}.  \end{equation} Then replacing
$\phi$ by $r^{k}$ in equation (\ref{16}) above it comes

\begin{equation} \label{19} V = \frac{k(k + 1)}{r^{2}} \end{equation} and
eliminating $r$ in relation (\ref{19}) above by replacing $r$ by $\phi^{1/k}$
gives

\begin{equation} \label{20} V \propto k(k + 1) \phi^{-2/k} \end{equation}

\begin{equation} \label{21} U \propto \int V \phi \,d\phi = \frac{k^{2}}{2}
\frac{k + 1}{k - 1} \phi^{2(1-1/k)}.  \end{equation} Therefore, relation
(\ref{21}) together with relation (\ref{17}) gives

\begin{equation} \label{22} k = - \frac{1}{\frac{p}{2} - 1}.  \end{equation} It
is natural to expect the exponent $p$ to be a positive integer.  Particularly,
we may consider the special case where this exponent is even.  Therefore, as $p
\geq 3$, it comes $p = 2(n + 1)$ where the integer $n$ runs from $1$ to
$\infty$.  This yields $\phi \propto r^{-1/n}$ and then the following potential

\begin{equation} \label{23} V_{\phi} = K\,r^{-1/n} \end{equation} where $K$ is
a real constant the magnitude of which depends on the strength of the
interaction of the scalar field $\phi$ with the ordinary matter in a given
region of spacetime.

\section{Application to the rotational curves of spiral galaxies} If a long
range scalar field such that the one considered in this paper does exist, then
as a dark matter candidate it may influence significantly the dynamics in a
spiral galaxy and thus modify the shape of its rotation curve.  Mostly, this
may be done without the need of a great amount of dark matter of this kind.
Indeed, setting $m^{(\phi)}_{dark}(r) = f \frac{dV_{\phi}}{dr}
\frac{r^{2}}{G}$, equation (\ref{9}) takes the form

\begin{equation} \label{24} \frac{dv_{r}}{dt} - \frac{v_{\theta}^{2}}{r} = - G
\frac{m(r) + m^{\phi}_{dark}(r)}{r^{2}} \end{equation} For $K < 0$, the
derivative $\frac{dV_{\phi}}{dr} = - \frac{K/n}{r^{1+1/n}}$ is positive and
thence $m^{(\phi)}_{dark}(r) = f \frac{\mid K \mid}{nG} r^{1-1/n}$ mimics a
dark matter mass profile which may be important though the energy corresponding
to the potential of the scalar field, $U(\phi)$, is in fact rather negligible
with respect to the real matter mass $m(r)$.  It is worth noticing that our
interpretation of $m^{(\phi)}_{dark}(r)$ as a hidden mass term involves that
the integer $n$ should run from $2$ to $\infty$ (instead of $1$ to $\infty$)
because one will require the positivity of the derivative
$\frac{dm^{(\phi)}_{dark}(r)}{dr}$ in addition to the positivity of
$m^{(\phi)}_{dark}(r)$ that is $\frac{dm^{(\phi)}_{dark}(r)}{dr} = \frac{n -
1}{n^{2}} f \frac{\mid K \mid}{G} r^{-1/n} > 0$ everywhere within a galaxy
(except at the boundary).  In this section, we wish to prove the validity of
relation (\ref{12}) on the basis of experimental data available on the rotation
curves of spiral galaxies.  To test relation (\ref{12}), it is better to
rewrite it as follows

\begin{equation} \label{25} \ln{(rv_{\theta})} = \ln{J} +
\frac{V_{\phi}}{c^{2}} \end{equation} and plot $\ln{(rv_{\theta})}$ versus $r$
for each rotation curve, where $r$ is the distance to the centre.  More
precisely we use the least-squares fit and search the coefficients $a$ and $b$
such that $\ln{(rv_{\theta})} = ar^{-1/n} + b$, where $a = \frac{K}{c^{2}}$ and
$b = \ln{J}$.  It turns out that the potential $V_{\phi}$ is indeed well
represented by a function of the form $K\,r^{-1/n}$, the exponent $n$ being the
integer that leads in absolute value to the highest correlation coefficient,
$R$, for a given rotation curve.  Consequently, the coefficient $a$ should be
negative in as much as the interpretation of $m^{(\phi)}_{dark}(r)$ as a
missing mass term holds.  Table $1$ below summarizes the numerical results
obtained for twenty three spiral galaxies including three dwarf galaxies (DDO
170, NGC 3109 and the dwarf "regular" UGC 2259) and three giant low surface
brightness disk galaxies (Malin 1, NGC 7589 and F 568-6).  Clearly, the
coefficient $a$ is always negative whereas $b$ is always positive.

\section{Fits to individual rotation curves}

The following figures show that the rotation curves of spiral galaxies
(covering a wide range of mass and size) can be reproduced essentially by a
potential, $V_{\phi}$, resulting from the interaction of a long range scalar
field with the usual galactic matter in as much as this interaction, like
gravity, respects the weak equivalence principle.  It is found that fairly good
fits can be obtained in the first approximation without the need to take into
account the photometric properties of the galaxies.\\\\

\begin{tabular}{|c|c|c|c|c|} \hline \emph{name [reference]} & \emph{n} &
\emph{$R^{2}$} & \emph{a} & \emph{b} \\ \hline DDO 170 \cite{21} & 3 &
0.999655088 & -7.23 & 9.8 \\ \hline DDO 170* \cite{21} & 3 & 0.99970403 & -7.1
& 9.7 \\ \hline NGC 3109 \cite{22} & 10 & 0.998736961 & -30.2 & 26.29 \\ \hline
NGC 3109* \cite{22} & 10 & 0.998784224 & -30.17 & 26.3 \\ \hline UGC 2259
\cite{23} & 7 & 0.999719756 & -15.14 & 16.89 \\ \hline Malin 1 \cite{24} & 3 &
0.999950275 & -11.67 & 12.33 \\ \hline NGC 7589 \cite{24} & 9 & 0.99981229 &
-16.11 & 19.72 \\ \hline F 568-6 \cite{24} & 9 & 0.999709336 & -17.67 & 20.97
\\ \hline NGC 4419 \cite{25} & 6 & 0.999743147 & -8.645 & 13.486 \\ \hline NGC
1035 \cite{25} & 5 & 0.99941517 & -8.4 & 12.51 \\ \hline NGC 1325 \cite{25} &
13 & 0.999646177 & -18.94 & 23.3 \\ \hline NGC 4062 \cite{26} & 18 &
0.999703876 & -24.49 & 29 \\ \hline NGC 2742 \cite{25} & 15 & 0.999726171 &
-20.7 & 25.2 \\ \hline NGC 3067 \cite{25} & 5 & 0.99959742 & -7.58 & 12.04 \\
\hline NGC 247 \cite{27} & 11 & 0.998423849 & -25.3 & 25.16 \\ \hline NGC 3198
\cite{28} & 5 & 0.999638676 & -6.716 & 11.52 \\ \hline UGC 12810 \cite{25} & 11
& 0.999759876 & -16.57 & 20.99 \\ \hline NGC 4051 \cite{29} & 8 & 0.998352026 &
-12.56 & 16.61 \\ \hline UGC 3691 \cite{26} & 7 & 0.999209314 & -12.57 & 16.14
\\ \hline NGC 3593 \cite{25} & 2 & 0.999652662 & -2.235 & 6.91 \\ \hline NGC
2639 \cite{25} & 2 & 0.998720197 & -5.77 & 9.92 \\ \hline NGC 4378 \cite{25} &
6 & 0.999521022 & -7.79 & 13.34 \\ \hline NGC 4448 \cite{25} & 4 & 0.998699541
& -6.306 & 11.073 \\ \hline NGC 7606 \cite{25} & 3 & 0.998640104 & -7.4 & 11.3
\\ \hline NGC 4402 \cite{30} & 13 & 0.99992101 & -27.85 & 29.22 \\ \hline
\end{tabular}\\\\{\bf Tab.  1.}  The exponent $n$, the square of the
correlation coefficient, $R^{2}$, and the parameters $a$ and $b$ of the
least-squares fit to the rotation curves of $23$ spirals are tabulated.  The
asterisk points out that the parameters presented are those of the final
rotation curve obtained after a correction for asymmetric drift (caused by a
pressure parameter) has been applied to the individual velocity points.  As
expected, the correlation coefficient increases when the correction for
asymmetric drift is applied.\\

\section{Concluding remarks} The work which has been presented in this paper
suggests the existence of a neutral scalar particle which may contribute
significantly to the rotation curves of spiral galaxies.  The results involved
by this proposal are found to be in good agreement with the available
observational data.  However, we have assumed that the referential attached, to
the dynamical centre of a given galaxy is galilean.  But, spiral galaxies are
disks in differential rotation.  So, a small correction may be necessary to fit
with more accuracy some peculiar rotation curves.  Besides, we have fitted the
rotation curves of $23$ spiral galaxies without the need of photometric data
since in our study the main contribution is brougt by the scalar field.
Nevertheless, this is not necessary problematical because it just brings one
more support to the fact that there exists a strong correlation between the
spatial distribution of the luminous matter and the spatial distribution of the
dark matter in spiral galaxies.

\end{document}